\documentclass[11pt,a4paper]{article}

\usepackage[T1]{fontenc}
\usepackage[utf8]{inputenc}
\usepackage{amsmath}
\usepackage{amssymb}
\usepackage{graphicx}
\DeclareGraphicsExtensions{.pdf,.eps,.png}
\usepackage{booktabs}
\usepackage[colorlinks=true,linkcolor=blue,citecolor=blue,urlcolor=blue]{hyperref}
\usepackage[margin=2.4cm]{geometry}

\newcommand{\dm}{\Delta m^2}
\newcommand{\Pmue}{P_{\mu e}}
\newcommand{\Pbar}{\overline{P}_{\mu e}}
\newcommand{\etaf}{f(E)}
\newcommand{\Mpl}{M_{\mathrm{Pl}}}
\newcommand{\cL}{c_L}

\title{\bfseries An energy-dependent rescaling of the neutrino
propagation Hamiltonian:\\ Standard Model Extension identification,
matter sector, and long-baseline degeneracy}

\author{N.~Razzaghi\\
\small Department of Physics, QA.C., Islamic Azad University, Qazvin, Iran\\
\normalsize \texttt{n.razzaghi@iau.ac.ir}}
\date{November 2026}

\begin{document}
\maketitle

\begin{abstract}
We study the consequences of multiplying the neutrino propagation
Hamiltonian by a real, energy-dependent scalar,
$H_\eta(E)=f(E)H_0(E)$ with $f(E)=1-\eta_0(E/E_0)^2$, a
minimal-length ansatz. Its vacuum part is a dimension-four,
CPT-even Lorentz-violating term of the Standard Model Extension
with coefficient $(\cL)_{ii}=-\eta_0(m_i^2-m_1^2)/(2E_0^2)$,
diagonal in the mass basis and fixed by the mass matrix, hence
removable by a reparametrization. The matter-potential rescaling
has dimension-five behaviour and is CPT-odd. For
$f=1-\eta_0(E/E_0)^p$ the vacuum term behaves as a dimension $p+2$
operator and the matter term as $p+3$, so only $p=2$ gives a
marginal vacuum operator; $p=0$ is absorbed into the measured mass
splittings, and non-integer $p$ admits no local operator. For
$\eta_0=0.02$ at $E_0=2.5$~GeV we find
$|(\cL)_{33}|\simeq4\times10^{-24}$, comparable to present
atmospheric bounds, whereas a Planck-suppressed deformation with
$\beta_0=\mathcal{O}(1)$ gives $3.2\times10^{-58}$. At constant
density it is equivalent to $L\to f(E)L$: mixing angles and the
Dirac phase are unchanged, the CP asymmetry is reparametrized
rather than generated, and the leading observable is a rescaling
of $\dm_{31}$. At $1297$~km a re-fit of $\dm_{31}$ alone absorbs
$23$~per cent of the squared probability difference, and a
four-parameter re-fit absorbs $41.5$~per cent. No statistical
analysis is performed; degeneracy means a distance in probability
space, not a confidence interval.
\end{abstract}

\section{Introduction}

Long baseline neutrino oscillation experiments are sensitive to
small modifications of the propagation Hamiltonian. Because the
oscillation phase grows with baseline and with energy, a deformation
that is negligible in a short baseline experiment can become
comparable to the leading atmospheric phase at a baseline of order a
thousand kilometres. This observation has motivated a large body of
work on minimal length scenarios, in which a deformed canonical
algebra modifies the dispersion relation of each mass eigenstate and
therefore, in principle, the oscillation pattern
\cite{Kempf:1994ek,Ali:2009zq,Das:2008kaa,BanerjeeDey,Ettefaghi,Gialamas}.

The purpose of this paper is narrow. We do not propose a new model. We
take a specific, frequently used phenomenological ansatz, namely a
multiplicative rescaling of the propagation Hamiltonian by a real
scalar that depends on energy as $E^2$. The defining feature of this
ansatz is that all mass squared differences are multiplied by one and
the same energy-dependent factor, so that the mixing matrix and the
Dirac phase are untouched. Deformations with exactly this structure
have been obtained from a minimal length, with a Gaussian rather than a
power law energy dependence, in Ref.~\cite{Sprenger:2011}, and the
power law version is used as a phenomenological input in
Ref.~\cite{Gialamas}. Other treatments of minimal length effects in
oscillations, in which the deformation is not a common factor but acts
through wave packet decoherence or through a state dependent
correction, are discussed in Refs.~\cite{Ettefaghi,BanerjeeDey}. We ask
three questions that can be answered without additional assumptions.

\begin{enumerate}
\item What is this ansatz in the language of the Standard Model
Extension (SME) \cite{Kostelecky:2003fs}? Is it a new form of
Lorentz violation or a familiar one?
\item What are its exact consequences for oscillation probabilities
at constant density, and which of those consequences are observable?
\item For a DUNE-like configuration \cite{DUNE}, how large is the
effect for a benchmark amplitude, and how much of it can be absorbed
by a re-fit of standard parameters?
\end{enumerate}

The answers, stated in advance, are the following. The vacuum part of
the ansatz can be written as a dimension four, CPT-even Lorentz
violating term with a coefficient that is diagonal in the mass basis
and proportional to the mass squared differences. That coefficient is
not an independent parameter of the theory: it is fixed by the mass
matrix, it is equivalent to a running of the mass squared differences,
and it can be removed by a reparametrization of the standard
parameters, so it carries no independent physical content; the
identification is a rewriting, not a new source of Lorentz violation.
Its natural size in a Planck
suppressed scenario is of order $10^{-58}$, that is, more than thirty
orders of magnitude below the benchmark amplitude that is usually
chosen for illustration. At constant density the ansatz is not merely
similar to a baseline rescaling; it is exactly a baseline rescaling,
with the rescaling factor depending on energy. Consequently the effect
does not mimic a CP phase in any exact sense, and the leading term is
approximately degenerate with $\dm_{31}$ rather than with
$\delta_{\rm CP}$. The word approximately is quantified in Sec.~\ref{sec:mimic}:
at a single energy a re-fit of $\dm_{31}$
absorbs between $88.6$ and $98.8$ per cent of the squared probability
difference, but across the beam band it removes only about a quarter of
it.

The value of this exercise, if any, is that it converts a family of
exploratory statements about quantum gravity sensitivity into a single
sharp statement about a dimension four coefficient that is already
constrained by atmospheric data \cite{Abe:2015lv}. We regard that
conversion as the main contribution of the paper. Because the analysis
stays at the level of probabilities, we present it as a note rather
than as a sensitivity study; Sec.~\ref{sec:stats} states explicitly
what a likelihood analysis would require and why we do not attempt
one.

\subsection{Relation to previous work}

The literature on minimal length corrections to neutrino oscillations
is large, and we do not attempt a review. Four strands are relevant
here. The first is the derivation of an $E^2$ correction to the
oscillation phase from a deformed algebra
\cite{Kempf:1994ek,Ali:2009zq,Das:2008kaa,BanerjeeDey,Ettefaghi,
Gialamas}, with related treatments in which the minimal length enters
through a wave packet or through an extended gravity background
\cite{Sprenger:2011,Lambiase:2003,Buoninfante:2020}. The outcome of
that derivation is a correction to the dispersion relation of each
mass eigenstate, which is what Eqs.~\eqref{eq:mdr}--\eqref{eq:dHGUP}
summarise. The second is the systematic classification of Lorentz and
CPT violating terms in the neutrino sector
\cite{Kostelecky:2003fs,Kostelecky:2011,Kostelecky:2012,Diaz:2009}, together with the
long baseline and atmospheric bounds on their coefficients
\cite{Abe:2015lv,Katori:2006,MINOS:2012,T2K:2017,IceCube:2018}. The
third is the family of parametrisations in which the mass squared
differences are taken to run with energy, $\dm_{ij}\propto E^n$, which
is closely related to a CPT violating neutrino mass matrix
\cite{Barenboim:2009}. The fourth is the treatment of propagation
decoherence and of non-standard matter effects, which we do not
address at all.

A word is needed on the relation between the present analysis and
decoherence, since the two are frequently conflated. Decoherence
arises when different mass eigenstates lose their relative phase
coherence, for instance through interactions with the environment or
through a spread of propagation times; it drives the oscillation
probability towards a weighted average over the mass eigenstates and
is described in the SME language by the non-Hermitian, CPT-even
coefficients that appear in the general framework of
Refs.~\cite{Kostelecky:2003fs,Kostelecky:2011,Kostelecky:2012,Diaz:2009}. The
rescaling studied here is the opposite kind of modification: it is a
coherent, Hermitian deformation that multiplies the whole Hamiltonian
by a real scalar, so that no coherence is lost, the oscillation
amplitude is preserved, and the only effect is a common rescaling of
the oscillation phases. The two therefore have different experimental
signatures: decoherence suppresses the amplitude at large $L/E$,
whereas a coherent rescaling leaves the amplitude intact and shifts
the phase. A minimal length scenario can in principle generate both,
and disentangling them requires an analysis that we do not attempt
here.

Two entries in the first two strands deserve a more specific comment,
because the operator that we discuss is a special case of a family
that has been written down before. In the general notation of the SME
the neutrino Hamiltonian may be expanded as
$\delta h_{ij}(E)=\sum_n \delta h^{(n)}_{ij}E^n$, and a term linear in
$E$ with a flavour non-diagonal coefficient is precisely the $d=4$
Lorentz violating operator of Eq.~\eqref{eq:cL}; our observation is
that the particular flavour structure
$(\cL)_{ij}\propto(m_i^2-m_j^2)\delta_{ij}$, i.e.\ a coefficient
diagonal in the mass basis and fixed by the mass squared differences,
is what a multiplicative rescaling of the whole Hamiltonian produces.
The same family contains the frequently considered parametrisation in
which the mass squared differences themselves are taken to run as
$\dm_{ij}\propto E^n$; that is a rescaling of the vacuum term alone,
and it differs from the present ansatz in that the matter potential is
left untouched. The two coincide only in vacuum, or when the matter
term is rescaled in the same way, which is the additional assumption
of Eq.~\eqref{eq:Vsync}. We are not claiming a new operator, but a
specific identification, which we have not found stated in this form,
of the operator that the multiplicative ansatz corresponds to, together
with its numerical consequence. The identification is in part implicit
in the literature on $\dm_{ij}\propto E^n$: what we add is the
statement that the coefficient is fixed by the mass matrix and is
therefore not an independent parameter.

The specific point that we have not found stated explicitly in this
form is the following. Once the correction is written as a
multiplicative factor on the full Hamiltonian, the vacuum part can be
written as a dimension four SME coefficient, with no expansion and
no approximation, and its size is fixed by the mass squared
differences rather than by an independent parameter. It is, however,
not an additional source of Lorentz violation: because the coefficient
is proportional to the mass matrix, it is degenerate with a shift of
the mass squared differences, and a reparametrization of the standard
parameters removes it. It is in this sense that the comparison with
the atmospheric bounds of Ref.~\cite{Abe:2015lv} can be made at the
level of a single number, and it is in this sense that the
gap of more than thirty orders of magnitude between the benchmark
amplitude and the Planck suppressed value is a statement about
coefficients rather than about probabilities. The exploratory
statements that we are commenting on are not wrong in their algebra;
they are statements about a coefficient that is already bounded, and
that in any case does not add a free parameter to the standard
description.

The paper is organized as follows. Section~\ref{sec:ansatz} defines
the ansatz and states precisely what is and what is not assumed.
Section~\ref{sec:sme} performs the mapping onto the SME and compares
with existing bounds. Section~\ref{sec:exact} derives the exact
solution and lists its consequences.
Section~\ref{sec:bench} gives the numerical benchmark for a DUNE-like
configuration. Section~\ref{sec:mimic} quantifies how much of the
effect a re-fit of standard parameters can absorb.
Section~\ref{sec:stats} explains what a genuine sensitivity study
would require and why we do not claim one. Section~\ref{sec:concl}
concludes.

\section{The ansatz and its logical status}
\label{sec:ansatz}

\subsection{Definition}

At fixed neutrino energy $E$ we write
\begin{equation}
H_\eta(E)=f(E)\,H_0(E),
\qquad
f(E)=1-\eta(E),
\qquad
\eta(E)=\eta_0\left(\frac{E}{E_0}\right)^2,
\label{eq:ansatz}
\end{equation}
where $H_0$ is the standard three flavour Hamiltonian in matter,
\begin{equation}
H_0(E)=\frac{1}{2E}\,U\,\mathrm{diag}(0,\dm_{21},\dm_{31})\,U^\dagger
+\mathrm{diag}(V_{CC},0,0),
\label{eq:H0}
\end{equation}
$U$ is the Pontecorvo, Maki, Nakagawa and Sakata matrix, and $V_{CC}$
is the charged current matter potential. The reference energy $E_0$
and the amplitude $\eta_0$ are not independent: only the product
$\eta_0/E_0^2$ enters Eq.~\eqref{eq:ansatz}. We keep both because
$E_0$ sets the energy at which the amplitude is quoted. We use the
convention $H=p-E$, so that a positive $\eta_0$ corresponds to a
reduction of the effective Hamiltonian and $f(E)<1$ at $E>0$;
throughout we take $\eta_0>0$. Because the ansatz is multiplicative,
the sign of $\eta_0$ is not a convention: $f$ is the ratio
$H_\eta/H_0$, and reversing the sign of the Hamiltonian reverses both
factors and leaves $f$ unchanged. The sign of $\eta_0$ is therefore
the sign of the deformation relative to the standard Hamiltonian, and
it is physical.

The form $\eta\propto E^2$ is the one that follows from a
Planck suppressed correction to the dispersion relation. Writing
$\beta=\beta_0/\Mpl^2$ for the deformation parameter of the algebra,
one has $\eta(E)=\kappa\beta_0 E^2/\Mpl^2$ with $\kappa$ a model
dependent coefficient, so that
\begin{equation}
\eta_0=\kappa\beta_0\frac{E_0^2}{\Mpl^2}.
\label{eq:eta0}
\end{equation}
We use the reduced Planck energy $\Mpl=2.435\times10^{18}$~GeV. For
the minimal length algebra of Sec.~\ref{sec:gup} one finds
$\kappa=3c_{\rm GUP}/2$. With the standard normalization
$c_{\rm GUP}=1$, which we adopt throughout, this is $\kappa=3/2$, and
we use that value whenever a numerical number is needed.

\subsection{Relation to a deformed dispersion relation}
\label{sec:gup}

It is important to be explicit about the status of
Eq.~\eqref{eq:ansatz}, because the literature sometimes presents it as
a consequence of a deformed algebra. It is not, but the relation is
closer than is usually stated, and it is worth deriving carefully.

Starting from a deformed commutator of the Kempf, Mangano and Mann
form \cite{Kempf:1994ek},
\begin{equation}
[\hat{x}_a,\hat{p}_b]=i\hbar\left[\delta_{ab}
+\beta\left(\hat{p}^2\delta_{ab}+2\hat{p}_a\hat{p}_b\right)\right],
\label{eq:gup}
\end{equation}
the leading correction to the dispersion relation of a mass eigenstate
is
\begin{equation}
E_i^2\simeq p^2\left(1-c_{\rm GUP}\,\beta_0
\frac{p^2}{\Mpl^2}\right)+m_i^2 .
\label{eq:mdr}
\end{equation}
What enters the oscillation probability is the spatial phase
$\Phi_i=\int p_i\,dx$, so Eq.~\eqref{eq:mdr} must be inverted for the
momentum. Writing $\epsilon=c_{\rm GUP}\beta_0E^2/\Mpl^2$ and
$\mu=m_i^2/E^2$, both small, and solving
$E^2=p^2(1-\epsilon p^2/E^2)+m_i^2$ order by order gives
\begin{equation}
p_i\simeq E-\frac{m_i^2}{2E}
+\frac{1}{2}c_{\rm GUP}\,\beta_0\frac{E^3}{\Mpl^2}
-\frac{3}{4}c_{\rm GUP}\,\beta_0\frac{m_i^2E}{\Mpl^2}
-\frac{m_i^4}{8E^3}.
\label{eq:pinv}
\end{equation}
The first correction,
$\delta p_{\rm GUP}=\tfrac12 c_{\rm GUP}\beta_0E^3/\Mpl^2$, is
independent of the mass eigenstate index. Its contribution to the
evolution operator is a common phase
$\exp(-i\int\delta p_{\rm GUP}\,dx)$, which cancels identically in
$P_{\alpha\beta}=|S_{\beta\alpha}|^2$. The term
$-\tfrac34 c_{\rm GUP}\beta_0\,m_i^2E/\Mpl^2$, the fourth term on the
right hand side of Eq.~\eqref{eq:pinv}, is however \emph{not}
universal: it is proportional to $m_i^2$ and therefore survives. It
corresponds to a Hamiltonian correction
\begin{equation}
\delta H_i^{\rm GUP}
=-\frac{3}{4}c_{\rm GUP}\,\beta_0\frac{m_i^2}{\Mpl^2}\,E ,
\label{eq:dHGUP}
\end{equation}
which is exactly the structure of Eq.~\eqref{eq:cL} below, with a
fixed coefficient. A deformed algebra of the form \eqref{eq:gup}
therefore does generate the vacuum part of the ansatz, with
$\kappa=3c_{\rm GUP}/2$, but it does not generate the
synchronous rescaling of the matter potential that we are forced to
adopt in Sec.~\ref{sec:exact}. We return to this point in
Sec.~\ref{sec:sme}. A word of caution is in order about the order to
which the expansion has been carried out. Equation~\eqref{eq:pinv} is
an expansion in two small parameters, $\epsilon=c_{\rm GUP}\beta_0
E^2/\Mpl^2$ and $\mu=m_i^2/E^2$, and it retains the leading term in
$\epsilon$ and the first correction in $\mu$. Terms of order
$\epsilon^2$, and in particular the
$\mathcal{O}(c_{\rm GUP}^2\beta_0^2E^5/\Mpl^4)$ term, have been
dropped from Eq.~\eqref{eq:pinv}; such a term would be smaller than the
retained $E^3/\Mpl^2$ term by $E^2/\Mpl^2\sim10^{-37}$ in the band of
interest, and being flavour universal it would cancel in the
probabilities in any case. The neglected $m_i^4/E^3$ term requires more care: it is
\emph{numerically} much larger than the deformation term
$m_i^2E/\Mpl^2$, by a factor $m_i^2\Mpl^2/E^4\sim10^{16}$ at
$E=1$~GeV, but it is a standard relativistic correction that is
present for $\beta_0=0$ and is not a Lorentz violating effect; it is
absorbed in the definition of the standard Hamiltonian and is
negligible in absolute terms, of relative order $(m_i/E)^2\sim
10^{-21}$. The statement that the leading \emph{deformation induced}
non-universal correction is the $m_i^2E$ term is therefore correct,
but it holds at leading order in both expansions, and a resummed
treatment would be needed if the deformation were not small. The
flavour universal $E^3$ term is kept in Eq.~\eqref{eq:pinv} only to
show that it cancels; it has no observable consequence.

The multiplicative form \eqref{eq:ansatz} is thus a phenomenological
hypothesis with a partial microscopic justification. We do not claim
more than that. The reason for studying it is that it is the simplest
deformation that preserves the probabilistic structure of the
evolution, and that it has a clean interpretation in the SME, as we
now show.

\section{Mapping onto the Standard Model Extension}
\label{sec:sme}

\subsection{The coefficient \texorpdfstring{$(\cL)_{ii}$}{(cL)ii}}

Consider the vacuum part of Eq.~\eqref{eq:H0} and multiply it by
$f(E)$ as in Eq.~\eqref{eq:ansatz}. The correction is
\begin{equation}
\delta H_{\rm vac}
=-\eta_0\frac{E^2}{E_0^2}\cdot
\frac{1}{2E}\,U\,\mathrm{diag}(0,\dm_{21},\dm_{31})\,U^\dagger
=-\frac{\eta_0 E}{2E_0^2}\,
U\,\mathrm{diag}(0,\dm_{21},\dm_{31})\,U^\dagger .
\label{eq:dH}
\end{equation}
The right hand side of Eq.~\eqref{eq:dH} is linear in $E$ and
independent of the baseline. In the SME the neutrino propagation
Hamiltonian contains, at dimension four and even under CPT, the term
\begin{equation}
(\delta H_{\rm SME})_{ij}=(\cL)_{ij}\,E ,
\label{eq:sme}
\end{equation}
where the coefficient is a matrix in flavour space for the isotropic
component. In the SME it is defined in the flavour basis, or
 equivalently as a tensor in the laboratory frame; the statement that
it is diagonal in the mass basis is a property of the present model,
not part of the definition, and the flavour basis coefficient follows
from it through the rotation
$(\cL)_{\rm fl}=U(\cL)_{\rm mass}U^\dagger$. Comparing
Eq.~\eqref{eq:dH} with Eq.~\eqref{eq:sme} gives immediately
\begin{equation}
(\cL)_{ii}=-\frac{\eta_0\,(m_i^2-m_1^2)}{2E_0^2}
=-\frac{\kappa\beta_0\,(m_i^2-m_1^2)}{2\Mpl^2},
\label{eq:cL}
\end{equation}
where in the second equality we used Eq.~\eqref{eq:eta0}. The
coefficient is diagonal in the mass basis, real, energy independent,
and proportional to the mass squared differences. It is not
proportional to the identity, so it is not removable by a common
phase. Had we written $m_i^2$ instead of $m_i^2-m_1^2$ the two
expressions would differ by the identity term
$-\eta_0m_1^2/(2E_0^2)$, which is unobservable, so only the traceless
part matters:
\begin{equation}
(\cL)_{ii}-\frac{1}{3}\mathrm{tr}\,\cL
=-\frac{\eta_0}{2E_0^2}\left(m_i^2-\frac{m_1^2+m_2^2+m_3^2}{3}\right).
\label{eq:traceless}
\end{equation}
which is the combination that enters the oscillation probabilities.

Equation~\eqref{eq:cL} is the central formal result of this paper, in
the limited sense explained below: it identifies the operator that the
ansatz corresponds to, not a new physical effect. Three remarks
follow.

First, the flavour structure of the ansatz is fixed: the coefficient is
proportional to the mass squared differences, that is, proportional to
the mass matrix in flavour space. This is the structure that a
coupling to a background field produces when the only flavour
symmetry breaking it feels is the neutrino mass matrix itself; a
coupling proportional to the identity in flavour space would not
break the degeneracy of the levels and would be unobservable. This is
a definite prediction of the ansatz, and it distinguishes it from a
generic Lorentz violating term, although, as we stress next, it also
makes the coefficient degenerate with the mass squared differences.
Note, however, that the comparison with experimental
bounds must be made in the same basis: the bounds of
Refs.~\cite{Abe:2015lv,Kostelecky:2011} are usually quoted for
flavour basis coefficients, so Eq.~\eqref{eq:cL} should be rotated
with $U$ before a quantitative comparison. Since the rotation mixes
the three entries, the bound on a single flavour basis component does
not translate into a bound on $\eta_0$ without a dedicated analysis.
The rotation involves products of elements of $U$ whose moduli lie
between $|U_{e3}|\simeq0.15$ and $|U_{\tau3}|\simeq0.72$, the latter
being the largest entry of the PMNS matrix; for comparison
$|U_{e2}|\simeq0.55$. The rotated entries therefore differ from the
mass basis ones by factors of order one and never by more than a factor
of a few. The order of magnitude
comparison made below is therefore unaffected, but a factor of two in
the quoted bound should not be read as significant.

Second, the ansatz is not a new energy regime. Its vacuum part is a
dimension four operator, and dimension four operators in the neutrino
sector are already constrained by atmospheric data, as discussed
below.

Third, the matter sector is different. The synchronous rescaling
$V_{CC}\to fV_{CC}$ of Eq.~\eqref{eq:Vsync} contributes
\begin{equation}
\delta H_{\rm mat}=-\eta_0\frac{E^2}{E_0^2}\,V_{CC},
\label{eq:dHmat}
\end{equation}
which is quadratic in $E$. With the normalisation
$\delta H=a^{(5)}E^2/\Mpl$ this has the energy behaviour of a
\emph{dimension five} operator, with coefficient
\begin{equation}
a^{(5)}=-\eta_0\frac{\Mpl\,V_{CC}}{E_0^2}\simeq-8.4\times10^{-7}
\qquad(\eta_0=0.02,\ E_0=2.5~{\rm GeV}),
\label{eq:a5}
\end{equation}
and with the opposite sign for antineutrinos, because $V_{CC}$
changes sign. It is therefore CPT-odd.\footnote{Throughout we use
$\hbar=c=1$ and the reduced Planck mass
$\Mpl=2.435\times10^{18}$~GeV. The normalisation
$\delta H=a^{(5)}E^2/\Mpl$ is the one in which $a^{(5)}$ is
dimensionless; other conventions in the literature differ by powers
of $\Mpl$, and we quote $a^{(5)}$ only to fix the order of magnitude.
For Eq.~\eqref{eq:a5} the arithmetic is
$a^{(5)}=-\eta_0\Mpl V_{CC}/E_0^2
=-0.02\times2.435\times10^{18}\times1.079\times10^{-22}/6.25
=-8.4\times10^{-7}$, with $V_{CC}=1.079\times10^{-22}$~GeV from
Eq.~\eqref{eq:VCC} and $E_0=2.5$~GeV. Using the unreduced Planck mass
$1.22\times10^{19}$~GeV instead would give $4.2\times10^{-6}$; the
difference is the factor $\sqrt{8\pi}$ and is the reason the convention
has to be stated.} It is also \emph{medium
dependent}: $V_{CC}$ is a property of the background through which the
neutrino propagates, not of the vacuum, so $a^{(5)}$ is not a
Lorentz violating coefficient of the Standard Model Extension in the
sense of Refs.~\cite{Kostelecky:2003fs,Kostelecky:2011}. The
dimension five operator of the SME with these quantum numbers is the
one that couples to the vacuum expectation value of a vector field;
the matter potential is not such an object, and Eq.~\eqref{eq:dHmat}
is therefore not a dimension five operator of the SME but a term with
the same energy behaviour. Forcing Eq.~\eqref{eq:dHmat}
into the dimension six form $\delta H=c^{(6)}E^3/\Mpl^2$ would require
an energy-dependent coefficient, $c^{(6)}=a^{(5)}\Mpl/E$, which
varies by a factor of seven across the band $0.6$--$4.0$~GeV, so no
constant $c^{(6)}$ exists. Equation~\eqref{eq:dHmat} is a separate
assumption, not implied by Eq.~\eqref{eq:mdr}, and it is the main
model dependence of the analysis. We flag it here and discuss its
consequences in Sec.~\ref{sec:exact}.

\subsection{General power laws and the dimension of the corresponding
operator}
\label{sec:tower}

The exponent $p=2$ in Eq.~\eqref{eq:ansatz} was fixed by the deformed
dispersion relation of Sec.~\ref{sec:gup}. It is worth asking what
happens for other powers, because the answer decides whether the
mapping of Sec.~\ref{sec:sme} is a property of the deformation or of
the exponent. We therefore replace Eq.~\eqref{eq:ansatz} by
\begin{equation}
f(E)=1-\eta_0\left(\frac{E}{E_0}\right)^{p},
\label{eq:fp}
\end{equation}
with $p$ a fixed real number, and repeat the two steps of
Sec.~\ref{sec:sme}. The vacuum correction becomes
\begin{equation}
\delta H_{\rm vac}
=-\frac{\eta_0}{2}\,\frac{E^{p-1}}{E_0^{p}}\,
U\,\mathrm{diag}(0,\dm_{21},\dm_{31})\,U^\dagger ,
\label{eq:dHp}
\end{equation}
which for $p=2$ is Eq.~\eqref{eq:dH}. In the normalisation
$\delta H^{(d)}=c^{(d)}E^{d-3}/\Mpl^{d-4}$, in which the coefficient
has mass dimension $4-d$ \cite{Kostelecky:2012}, the power $E^{p-1}$
corresponds to
\begin{equation}
d_{\rm vac}=p+2 ,
\qquad
c^{(d)}_{ii}=-\frac{\eta_0}{2}\,
\frac{(m_i^2-m_1^2)\,\Mpl^{\,d-4}}{E_0^{\,d-2}} ,
\label{eq:cd}
\end{equation}
so that $d_{\rm vac}=4$ and Eq.~\eqref{eq:cd} reduces to
Eq.~\eqref{eq:cL} only for $p=2$. The matter term of
Eq.~\eqref{eq:dHmat} generalises in the same way,
\begin{equation}
\delta H_{\rm mat}=-\eta_0\,\frac{E^{p}}{E_0^{p}}\,V_{CC}
\qquad\Longrightarrow\qquad
d_{\rm mat}=p+3 ,
\label{eq:dmat}
\end{equation}
with coefficient $a^{(d)}=-\eta_0\Mpl^{\,p-1}V_{CC}/E_0^{\,p}$, which
for $p=2$ is Eq.~\eqref{eq:a5}. The two dimensions therefore differ by
one for every $p$,
\begin{equation}
d_{\rm mat}=d_{\rm vac}+1 ,
\label{eq:ddiff}
\end{equation}
which is the general form of the statement, made in
Sec.~\ref{sec:sme}, that the matter sector sits one dimension higher
than the vacuum sector. Table~\ref{tab:tower} collects the
classification.

\begin{table}[t]
\centering
\caption{Classification of the power law of Eq.~\eqref{eq:fp}. The
vacuum term has the energy behaviour of a dimension $p+2$ operator and
the matter term of a dimension $p+3$ operator, for every $p$. The CPT
column is independent of $p$: the vacuum term is CPT-even and the
matter term CPT-odd for all $p$. Only $p=2$ gives a marginal, that is
renormalizable, vacuum operator.}
\label{tab:tower}
\begin{tabular}{lcccc l}
\toprule
$p$ & $d_{\rm vac}$ & $d_{\rm mat}$ & vacuum CPT & matter CPT
& vacuum operator \\
\midrule
$0$ & $2$ & $3$ & even & odd & relevant (absorbed) \\
$1$ & $3$ & $4$ & even & odd & relevant \\
$2$ & $4$ & $5$ & even & odd & marginal \\
$3$ & $5$ & $6$ & even & odd & irrelevant \\
$4$ & $6$ & $7$ & even & odd & irrelevant \\
\bottomrule
\end{tabular}
\end{table}

Three remarks follow.

First, $p=2$ is the unique power for which the vacuum term is
marginal. For $p>2$ the operator is irrelevant: its coefficient carries
a negative mass dimension, and the term is suppressed by powers of
$E/\Mpl$ relative to the standard Hamiltonian, so it is unobservable
for the same reason as the Planck suppressed case discussed in
Sec.~\ref{sec:sme}. For $p<2$ the operator is relevant, and it is
worth separating the two cases, because they are not alike. At $p=0$
Eq.~\eqref{eq:dHp} reduces exactly to a constant multiple of the
vacuum Hamiltonian, so it rescales the mass matrix and nothing else;
the correction is then absorbed into the measured $\dm_{ij}$ and is
invisible in oscillation data. For $0<p<2$ this is no longer true. The
correction is not proportional to $H_0^{\rm vac}$: at $p=1$, for
instance, Eq.~\eqref{eq:dHp} is
$-\eta_0(2E_0)^{-1}U\,{\rm diag}(0,\dm_{21},\dm_{31})U^\dagger$,
which is constant in $E$ and therefore a mass-like term with no
$(\cL)$ counterpart at its nominal dimension, rather than a rescaling
of the vacuum Hamiltonian. Such a term is in principle observable, and
it is constrained by oscillation data without reference to $\Mpl$; but
the constraint is on the combination that multiplies the mass matrix,
not on a Lorentz violating coefficient, so it does not translate into
a bound on $\beta_0$ in the way the $p=2$ case does. The exponent
$p=2$ is therefore the only one for which the vacuum term is at once
marginal, not absorbed, and identifiable with a coefficient of the
Standard Model Extension, which is a statement about the exponent
rather than about the model, and it explains why the identification of
Sec.~\ref{sec:sme} is not an accident of the choice.

Second, the CPT properties are independent of $p$, and this is worth
stating because the dimension label is often read as carrying them. A
real $f(E)$ multiplies the Hamiltonians of neutrinos and antineutrinos
by the same factor, so the vacuum correction of Eq.~\eqref{eq:dHp} is
CPT-even for every $p$. The matter correction of Eq.~\eqref{eq:dmat}
is proportional to $V_{CC}$, which changes sign under the replacement
of neutrinos by antineutrinos, so it is CPT-odd for every $p$. The
dimension in Table~\ref{tab:tower} classifies the energy scaling only.
The distinction matters at odd $p$: the vacuum term then has the energy
behaviour of an odd dimension operator while remaining CPT-even, and in
the operator tower of Ref.~\cite{Kostelecky:2012} the coefficient
$(\cL)_{ij}$ exists only at even $d$, the CPT-even coefficients at odd
$d$ being mass-like. For $p=1$, for instance, Eq.~\eqref{eq:dHp} is
constant in $E$ and is a mass term rather than a $(\cL)$ term. An odd
$p$ vacuum correction therefore has no $(\cL)$ counterpart at its
nominal dimension, and the dimension four identification of
Sec.~\ref{sec:sme} is specific to $p=2$ in a second, independent sense.

Third, a non-integer $p$ admits no local operator at all. A local term
in the Lagrangian produces integer powers of $E$ in the propagation
Hamiltonian, whereas $E^{p}$ with non-integer $p$ is not analytic at
$E=0$. The mapping of Sec.~\ref{sec:sme} is then empty, and the power
law of Eq.~\eqref{eq:fp} can only be read as a purely phenomenological
parametrisation, with no operator interpretation.

The physical content of this tower is negative, and it is the same
negative content as before. At constant density the equivalence
$L\to f(E)L$ holds for every $p$, so the whole tower is degenerate with
an energy-dependent re-fit of $\dm_{31}$ and of $V_{CC}$, and the CP
asymmetry is reparametrised, $A_\eta(E)=A_0(f(E)L)$, rather than
generated. The tower therefore adds no observable. What it adds is a
classification: it shows that the dimension four identification of
Eq.~\eqref{eq:cL} is the marginal member of a one parameter family, and
that the choice of exponent, not the choice of deformation, is what
makes the ansatz look like a renormalizable Lorentz violating operator.

\subsection{Numerical size and comparison with bounds}

For the benchmark used throughout this paper, $\eta_0=0.02$ and
$E_0=2.5$~GeV, Eq.~\eqref{eq:cL} with $m_3^2-m_1^2=2.521\times
10^{-3}$~eV$^2$ gives
\begin{equation}
|(\cL)_{33}|\simeq 4.0\times10^{-24}.
\label{eq:cLnum}
\end{equation}
The same benchmark corresponds, through Eq.~\eqref{eq:eta0} with
$\kappa=3/2$, to $\beta_0^{\rm match}\simeq1.3\times10^{34}$.
Conversely, a genuinely Planck suppressed deformation with
$\beta_0=1$ gives
\begin{equation}
|(\cL)_{33}|\simeq 3.2\times10^{-58},
\qquad
\eta(E_0)\simeq 1.6\times10^{-36}.
\label{eq:cLplanck}
\end{equation}

The numbers in Eqs.~\eqref{eq:cLnum} and \eqref{eq:cLplanck} differ by
more than thirty orders of magnitude, and this difference is the whole
content of the frequently repeated statement that a DUNE-like
experiment cannot probe Planck suppressed minimal length effects
through this channel. The same conclusion is reached, by a different
route, in Ref.~\cite{Gialamas}: there the bound
$\beta_0\lesssim10^{35}$ is quoted from the projected sensitivity of
a combined Hyper-K and JUNO solar flux analysis to non-standard
interactions, $|\varepsilon|\lesssim0.1$, rather than from an
existing measurement. One word of warning about the symbols is in
order, since the two papers use the same letter for different
quantities. The parameter of the algebra is called $\beta_0$ in both
papers, and Eq.~\eqref{eq:eta0} relates it to our amplitude $\eta_0$;
the bound $\beta_0\lesssim10^{35}$ is a bound on the algebra
parameter, not on $\eta_0$. Taken at face value, and with
$\kappa=3/2$, $E_0=2.5$~GeV and $\Mpl=2.435\times10^{18}$~GeV, it
corresponds to $\eta_0\lesssim0.16$, so the benchmark $\eta_0=0.02$
used here lies
a factor of eight below the projected reach quoted there. The two
statements are of a different nature, however, and this is why the
comparison in this paragraph is made between coefficients rather
than between these parameters: the bound on $\beta_0$ is a projection
for a future solar flux analysis, whereas the constraint on
$(\cL)_{ii}$ discussed below is derived from existing atmospheric
data. The coefficient is the quantity that appears in the SME
Lagrangian, and therefore the quantity for which experimental bounds
exist.

Atmospheric neutrino data constrain the isotropic components of
$(\cL)_{ij}$ at the level of a few times $10^{-24}$ to $10^{-23}$,
depending on the component and on the analysis
\cite{Abe:2015lv,Kostelecky:2011}. The benchmark of
Eq.~\eqref{eq:cLnum} therefore sits at the edge of the existing
sensitivity rather than many orders of magnitude beyond it. This is a
more informative statement than the one usually made, and it is the
reason why we regard the channel as worth quantifying: a dedicated
analysis at a long baseline experiment would improve a bound on a
dimension four coefficient by a modest factor, not discover quantum
gravity.

We stress that Eq.~\eqref{eq:cLnum} is not a bound and not a
prediction. It is the coefficient that corresponds to the illustrative
amplitude $\eta_0=0.02$. The physically motivated value is
Eq.~\eqref{eq:cLplanck}.

\section{Exact solution}
\label{sec:exact}

\subsection{The evolution operator}

At constant density the Hamiltonian does not depend on the position,
and the evolution operator over a baseline $L$ is
\begin{equation}
S_\eta(E,L)=\exp\left(-i\,H_\eta(E)\,L\right)
=\exp\left(-i\,\etaf\,H_0(E)\,L\right)
=\exp\left(-i\,H_0(E)\,L_{\rm eff}\right),
\label{eq:Seff}
\end{equation}
with
\begin{equation}
L_{\rm eff}(E)=\etaf\,L .
\label{eq:Leff}
\end{equation}
Equation~\eqref{eq:Seff} is exact. It is not an expansion and it does
not require $H_0$ to be diagonal: it uses only that $f(E)$ is a real
scalar at fixed energy, so that it commutes with $H_0(E)$ and can be
absorbed into the time variable.

If the density is not constant, the situation is different, and the
difference matters. The evolution operator is the ordered exponential
\begin{equation}
S_\eta=\mathcal{T}\exp\left(-i\int_0^L\! H_\eta(x)\,dx\right)
=\mathcal{T}\exp\left(-i\int_0^L\! \etaf H_0(x)\,dx\right),
\label{eq:Svar}
\end{equation}
and since $f(E)$ is a c-number that does not depend on $x$ it can be
pulled out of the ordered product. What is left is not the standard
evolution operator at a shifted baseline, however, because the
position variable is also rescaled. Writing $x'=f x$,
\begin{equation}
S_\eta=\mathcal{T}\exp\left(-i\int_0^{\etaf L}
H_0(x'/\etaf)\,dx'\right),
\label{eq:Sstretch}
\end{equation}
so the ansatz is equivalent to a shorter baseline \emph{and} a
profile stretched by $1/f$. The two effects coincide only when the
profile is constant, or more generally when it is self similar under
$x\to x/f$ over the relevant range. For a varying profile a
discontinuity located at $x_d$ along the path is displaced by
$(1-f)\,x_d$, not by $(1-f)L$; the latter is the special case
$x_d=L$. For $f=0.949$ ($E=4$~GeV) a discontinuity in the middle of a
$1297$~km baseline is therefore displaced by about $33$~km. The
Moho, taken at a depth of about $33$~km,\footnote{The Moho lies
between roughly $25$~km beneath the oceans and $40$--$70$~km beneath
the continents. We use the value met by this particular chord; the
crossing points quoted shift accordingly if a different depth is
assumed, and the conclusion, which is that the displacement is set by
the position along the path, does not.} is met by such a chord at
$x_d\simeq 615$~km and $683$~km, so its two crossings are displaced
by about $31$~km and $35$~km, not by a small amount: the displacement
is set by the position along the path, not by the depth. For the
Earth the Moho is the only structural discontinuity that lies on the
path. With $R_\oplus=6371$~km and a core radius of $3480$~km the
chord of a $1297$~km baseline reaches a maximum depth of $33.1$~km,
its closest approach to the centre being $6338$~km, and the shortest
baseline that crosses the core is $10673$~km. A $1297$~km beam
therefore never samples the mantle--core discontinuity, and the
statement $x_d=L$, which would give the $66$~km figure, is not
realised. The resulting change in the oscillation probability is small
for a smooth profile, of relative order $10^{-3}$ on the band that we
consider, but it is not zero, and Eq.~\eqref{eq:Sstretch} is the
correct statement. We have verified it numerically: for a two step
profile the evolution operator of Eq.~\eqref{eq:Sstretch} agrees with
the direct integration of Eq.~\eqref{eq:Svar} to better than
$10^{-11}$, whereas the naive baseline rescaling at fixed profile
differs from it at the $10^{-3}$ level. This comparison is a test of
the correctness of the change of variables, not a physical statement:
Eq.~\eqref{eq:Sstretch} is an identity, so the two sides must agree up
to the accuracy of the integration, and the residual $10^{-11}$
reflects the discretisation of the numerical evolution rather than any
dynamical effect.

It follows that the ansatz introduces no new interference phenomenon
and no new level crossing: it only shortens the optical path, and
stretches the medium along it. In the standard definition the
adiabaticity parameter is
$\gamma=\Delta/(2E\,|d\theta_m/dx|)$, the ratio of the level splitting
to the rate of change of the matter mixing angle, and it is this
quantity that governs the transition probability. Under $x\to x/f$ the
splitting is scaled by $f$ while the gradient $d\theta_m/dx$ is left
unchanged, so $\gamma$ acquires a factor $f$ exactly; if the profile
is stretched as well, the gradient is reduced by a further factor $f$
and the net factor is $f^2$. Either way the adiabaticity of the
propagation is modified by the same amount as the baseline. For $f$
close to one, which is the only regime in which the ansatz is under
control, the modification is a small correction.
It is not, however, a statement that nothing changes.

The consequences follow immediately and are worth listing, because
several of them contradict statements that appear in the exploratory
literature.

\begin{enumerate}
\item \emph{Mixing angles are unchanged.} The eigenvectors of
$H_\eta$ coincide with those of $H_0$, so the matter mixing angles
$\theta_{12}^m$, $\theta_{13}^m$ and $\theta_{23}^m$ are unaffected at
each energy.
\item \emph{All mass squared differences are multiplied by the same
factor.} The eigenvalues of $H_\eta$ are $f$ times those of $H_0$, so
the effective mass squared differences obey
$(\Delta m^2)_{ij}^{\rm eff}=f\,(\Delta m^2)_{ij}^{\rm matter}$. In
particular the atmospheric and solar splittings are rescaled by the
same amount.
\item \emph{The oscillation phase is rescaled.} Since the phase is the
product of the eigenvalue difference and the baseline, the rescaling
can be attributed either to the eigenvalues or to the baseline, and
Eq.~\eqref{eq:Seff} shows that the two descriptions are identical.
\item \emph{The CP asymmetry is not generated.} Because the same
factor multiplies the neutrino and the antineutrino Hamiltonians at a
given energy, the asymmetry
$A_{\rm CP}=(P-\bar P)/(P+\bar P)$ satisfies the exact relation
$A_{\rm CP}(E)=A_{\rm CP}^0\big(f(E)L\big)$ at constant density,
where $A_{\rm CP}^0$ is the standard asymmetry evaluated at the
shifted baseline. The relation holds only in that approximation: it
follows from the exact equivalence $L\to f(E)L$, which requires a
constant density, and for a varying profile it becomes the stretched
profile of Eq.~\eqref{eq:Sstretch}, in which $f$ enters through the
position along the path rather than through the baseline alone. The ansatz does not create a CP violating effect;
it relocates the standard one. This is a useful diagnostic: any
claimed enhancement of the CP violation in this class of models must
come from somewhere other than the multiplicative factor. We have
checked the relation numerically at $E=1.0$, $2.5$ and $4.0$~GeV and
for $\delta_{\rm CP}=0$ and $\pi/2$; it holds to better than
$10^{-14}$ in all six cases.
\item \emph{The MSW resonance is \emph{not} displaced.} Under the
synchronous scaling of Eq.~\eqref{eq:Vsync} the eigenvalues of
$H_\eta$ are $f$ times those of $H_0$ and the eigenvectors are
identical, so the energy at which the matter mixing angle
$\theta_{13}^m$ is maximal is the same as in the standard case. A
direct numerical scan of the three-level Hamiltonian between $0.5$ and
$15$~GeV confirms this. At $0.5$~GeV the two level spacings are
$1.03\times10^{-13}$ and $2.38\times10^{-12}$~eV; at $15$~GeV they
are $7.58\times10^{-14}$ and $3.76\times10^{-14}$~eV. Neither spacing
decreases monotonically. The lower one has a shallow local maximum of
$1.01\times10^{-13}$~eV at about $4.7$~GeV, having fallen to
$9.73\times10^{-14}$~eV at $1$~GeV, and reaches its minimum at
$15$~GeV; the upper one peaks near $0.3$~GeV and is minimal near
$12$~GeV. The overall trend is a decrease with energy, as expected from
the growth of the matter potential, but it is not monotonic, and no
two levels cross anywhere in the range. The decrease is present with
or without the factor $f$. The absence of a crossing in this band is
not in conflict with the vanishing of $f$ at
$E_{\rm max}=17.68$~GeV discussed below: at $E_{\rm max}$ the three
levels become exactly degenerate because the whole Hamiltonian is
multiplied by zero, which is a property of the truncation of $f(E)$
rather than a level crossing inside the physically accessible band.
For the parameters of Eq.~\eqref{eq:params} the resonance lies at
$E_{\rm res}=\dm_{31}\cos2\theta_{13}/(2V_{CC})\simeq11.2$~GeV, that
is, outside the beam band of a DUNE-like experiment, so the resonance
plays no role in the analysis of Sec.~\ref{sec:bench}. We emphasize
this because the opposite conclusion is easy to reach by applying the
resonance condition to the vacuum splitting alone; see
Sec.~\ref{sec:matter}.
\end{enumerate}

\subsection{The matter potential and the range of validity}
\label{sec:matter}

Equation~\eqref{eq:Seff} treats the full Hamiltonian, vacuum and
matter, as rescaled by the same factor. This is a choice, and it must
be stated as such. If instead only the vacuum term were rescaled, then
at constant density the resonance condition would read
$V_{CC}=f\,\dm_{31}\cos2\theta_{13}/(2E)$ with $V_{CC}$ unscaled, and
the resonance would move. For the parameters of
Eq.~\eqref{eq:params} the solution of
$2EV_{CC}=f(E)\dm_{31}\cos2\theta_{13}$ is $E_{\rm res}\simeq8.5$~GeV,
a downward shift of about $23\%$ with respect to the standard value.
We therefore adopt the synchronous scaling of the matter potential,
\begin{equation}
V_{CC}\to \etaf\,V_{CC},
\label{eq:Vsync}
\end{equation}
as part of the definition of the ansatz. Equation~\eqref{eq:Vsync} is
an assumption, not a derived result. In the SME language of
Sec.~\ref{sec:sme} it corresponds to the CPT-odd, medium-dependent
term of Eq.~\eqref{eq:dHmat} with dimension five energy behaviour,
which is not implied by Eq.~\eqref{eq:cL}. A reader who prefers to keep the matter
potential fixed will find that the resonance is displaced as described
above, and that the numerical results of Sec.~\ref{sec:bench} change
accordingly. This is the assumption already flagged in
Sec.~\ref{sec:sme} as the main model dependence of the analysis.

A second, more serious limitation is the range of validity in energy.
The function $f(E)=1-\eta_0(E/E_0)^2$ vanishes at
\begin{equation}
E_{\rm max}=\frac{E_0}{\sqrt{\eta_0}}=17.68~\mathrm{GeV}
\qquad(\eta_0=0.02,\ E_0=2.5~\mathrm{GeV}),
\label{eq:Emax}
\end{equation}
and becomes negative above it. Requiring $f>0$ over the whole range in
which the ansatz could plausibly be applied, say up to $100$~GeV,
gives the bound $\eta_0<6.25\times10^{-4}$; the benchmark
$\eta_0=0.02$ used here for illustration therefore lies outside the
window of validity of the truncation above $17.7$~GeV, and no
statement in this paper depends on extrapolating it there. At $E=E_{\rm max}$ the evolution
operator is the identity and the appearance probability vanishes
exactly; for $E>E_{\rm max}$ the sign of the Hamiltonian is reversed,
which by the symmetry $P(L)=P(-L)$, valid at constant density for a
Hamiltonian that is linear in the oscillation phase, returns the
standard probability at $f=-1$, that is, at
$E=25$~GeV, provided that the standard parameter $\delta_{\rm CP}$
vanishes or $\pi$. For $\delta_{\rm CP}=0$ we have checked explicitly
that $P_\eta(f=-1;L)=P_0(L)$ at the $10^{-15}$ level, both for
neutrinos and for antineutrinos, the common value being
$1.32\times10^{-3}$ for the benchmark of Eq.~\eqref{eq:params}. For
non-zero $\delta_{\rm CP}$ the relation is
$P_\eta(f=-1;L)=P_0(-L)=P_0(L)^T$, which equals $P_0(L)$ only when the
$T$ violating interference term is absent. The numbers that we quote
below for illustration, $1.144\times10^{-3}$ against
$1.074\times10^{-3}$ for the benchmark of Eq.~\eqref{eq:params}, are
for $\delta_{\rm CP}=\pi/2$, where the difference is largest; they are
not a test of the $\delta_{\rm CP}=0$ statement, for which the two
sides agree exactly. The statement about the sign
reversal is therefore a statement about the Hamiltonian, not about
the probability. A model that predicts a vanishing appearance probability
at $17.7$~GeV is grossly excluded by atmospheric and long baseline
data. The ansatz must therefore be regarded as an expansion in
$(E/E_0)^2$ that is truncated at first order and used only for
$E\ll E_{\rm max}$. All numerical results of this paper refer to the
band $0.6$--$4.0$~GeV, where $f\in[0.949,0.999]$ and the truncation is
under control, and we do not extrapolate beyond it. This restriction
should be kept in mind whenever the ansatz is applied to atmospheric
or high energy astrophysical neutrinos, where $E\gg E_0$ and the
first order truncation is not justified.

\section{Numerical benchmark for a DUNE-like configuration}
\label{sec:bench}

\subsection{Parameters}

We use the normal mass ordering and the following representative
values, of the same order as the NuFIT 6.0 normal ordering best fit
\cite{NuFIT}:
\begin{align}
\sin^2\theta_{12}&=0.3088, &
\sin^2\theta_{13}&=0.02249, &
\sin^2\theta_{23}&=0.470, \nonumber\\
\dm_{21}&=7.537\times10^{-5}~\mathrm{eV}^2, &
\dm_{31}&=2.521\times10^{-3}~\mathrm{eV}^2, &
\delta_{\rm CP}&=0 .
\label{eq:params}
\end{align}
These are rounded values chosen for definiteness; we do not use the
exact global best fit, and none of our conclusions depends on the
difference. The one entry for which the difference is worth stating
explicitly is $\dm_{31}$: the NuFIT 6.0 normal ordering best fit is
$2.51\times10^{-3}$~eV$^2$, within about half a per cent of the value
used here.
That is smaller than the spread between the published global fits, and
far smaller than the $2.5\%$ shift of $\dm_{31}$ that the ansatz
itself induces at $\eta_0=0.02$, so it does not affect any number in
this paper; we keep $2.521\times10^{-3}$~eV$^2$ because it is the value
at which the benchmark probabilities of Sec.~\ref{sec:bench} were
computed. The value $\delta_{\rm CP}=0$ is an illustrative CP
conserving point. For the matter density we take
$\rho=2.827$~g~cm$^{-3}$ with electron fraction $Y_e=0.5$, which gives
\begin{equation}
N_e=8.512\times10^{23}~\mathrm{cm}^{-3},
\qquad
V_{CC}=1.079\times10^{-13}~\mathrm{eV},
\label{eq:VCC}
\end{equation}
and we take $L=1297$~km. Where the matter potential is allowed to
float in the fits of Sec.~\ref{sec:mimic} it is parametrised by the
ratio
\begin{equation}
s_V\equiv\frac{V_{\rm eff}}{V_{CC}},
\label{eq:sV}
\end{equation}
with $V_{CC}$ the value of Eq.~\eqref{eq:VCC}, so that $s_V=1$
reproduces the PREM motivated density used throughout. The value
quoted in Eq.~\eqref{eq:VCC} is the one implied by
$\rho=2.827$~g~cm$^{-3}$;
with the slightly larger $\rho=2.848$~g~cm$^{-3}$ that is sometimes
quoted for the same baseline one would obtain
$N_e=8.575\times10^{23}$~cm$^{-3}$, a $0.7\%$ difference which changes
$V_{CC}$ by the same relative amount and is absorbed by the
$2.5\%$ matter potential freedom already included in
Table~\ref{tab:fits}. The constant density treatment is a
simplification; the Preliminary Reference Earth Model \cite{PREM}
exhibits variations of order ten per cent along this baseline. As
shown in Sec.~\ref{sec:exact}, the constant density identity
\eqref{eq:Seff} is replaced by Eq.~\eqref{eq:Sstretch} for a varying
profile, so the constant density assumption affects the numerical
values but not the structure of the result. We return to this point
in Sec.~\ref{sec:stats}.

\subsection{Probabilities at the reference energy}

\begin{figure}[t]
\centering
\includegraphics[width=\textwidth]{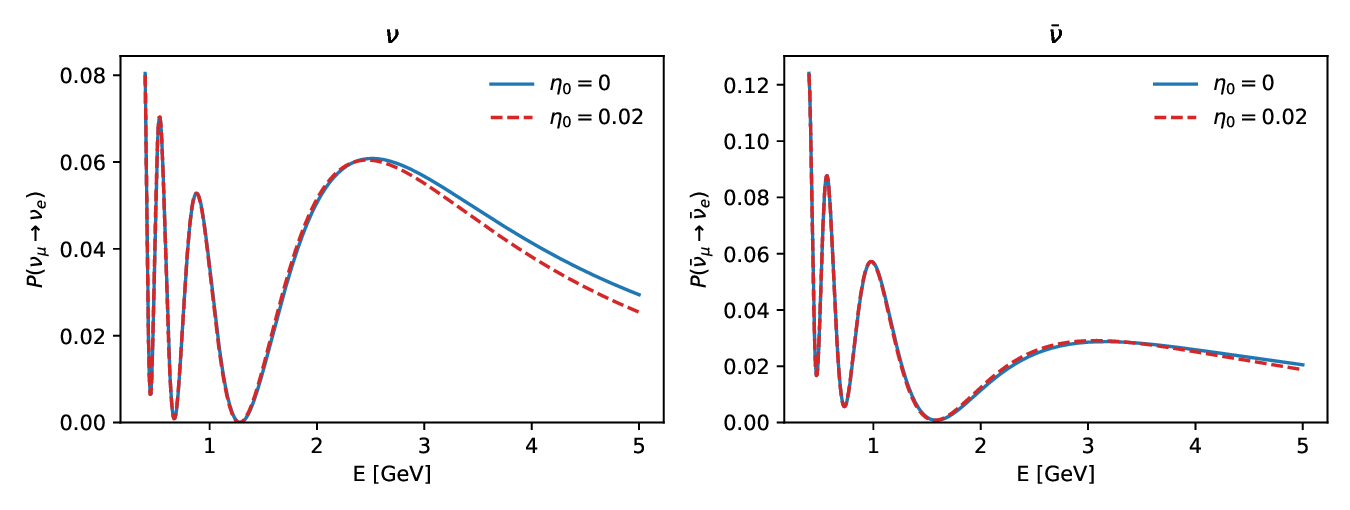}
\caption{Appearance probability $\Pmue$ as a function of energy for
$\eta_0=0$ (solid) and $\eta_0=0.02$ (dashed), for neutrinos (left)
and antineutrinos (right), with the parameters of
Eq.~\eqref{eq:params} and a baseline of $1297$~km.}
\label{fig:prob}
\end{figure}

At $E=2.5$~GeV, where $f=0.98$, the exact evolution operator of
Eq.~\eqref{eq:Seff} gives
\begin{equation}
\begin{aligned}
\Pmue^{0}&=0.060820, &
\Pmue^{\eta}&=0.060417,\\
\Pbar^{0}&=0.024323, &
\Pbar^{\eta}&=0.025400,
\end{aligned}
\label{eq:bench}
\end{equation}
for $\eta_0=0.02$; the full energy dependence is shown in
Fig.~\ref{fig:prob}. The relative changes are $-0.66\%$ for neutrinos and
$+4.4\%$ for antineutrinos. The sign difference is not a CP violating
effect: it follows from the fact that the neutrino and antineutrino
probabilities have different dependences on the oscillation phase, and
the same factor $f$ moves them in opposite directions where the
derivative with respect to the phase has opposite sign.

\subsection{Energy dependence of the response}

\begin{figure}[t]
\centering
\includegraphics[width=0.72\textwidth]{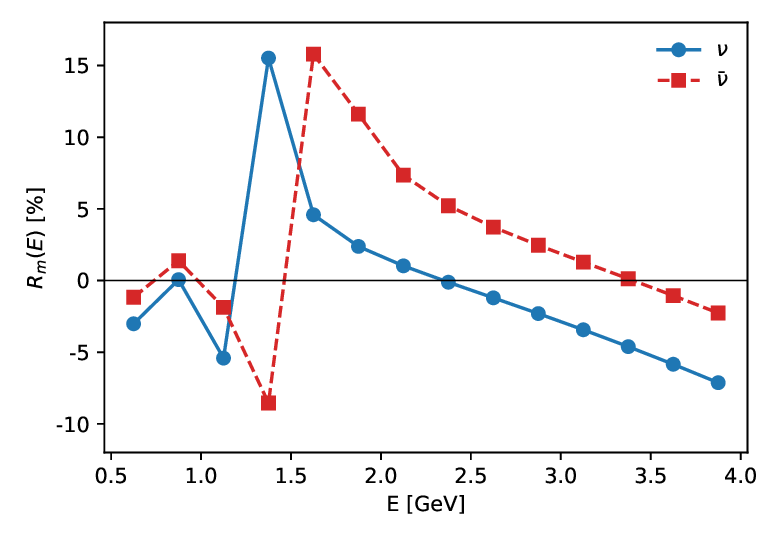}
\caption{Signed fractional response $R_m(E)$ of Eq.~\eqref{eq:R} on the
energy nodes, for neutrinos (circles) and antineutrinos (squares).}
\label{fig:response}
\end{figure}

The benchmark of Eq.~\eqref{eq:bench} is a single energy. What matters
for an experiment is the behaviour across the beam band. On the
fourteen nodes $E_i=(0.625+0.25\,i)$~GeV, $i=0,\ldots,13$, we define
the signed fractional response
\begin{equation}
R_m(E)=100\,\frac{P_m^{\eta}(E)-P_m^{0}(E)}{P_m^{0}(E)},
\label{eq:R}
\end{equation}
with $m=\nu,\bar\nu$. The results are collected in
Table~\ref{tab:response} and shown in Fig.~\ref{fig:response}.

\begin{table}[t]
\centering
\caption{Signed fractional response $R_m(E)$ in per cent, defined in
Eq.~\eqref{eq:R}, for $\eta_0=0.02$ and the parameters of
Eq.~\eqref{eq:params}. The extrema on the node grid are marked in
bold.}
\label{tab:response}
\begin{tabular}{ccc}
\toprule
$E$ [GeV] & $R_\nu$ [\%] & $R_{\bar\nu}$ [\%] \\
\midrule
0.625 & $-3.02$ & $-1.17$ \\
0.875 & $0.06$  & $1.38$  \\
1.125 & $-5.41$ & $-1.87$ \\
1.375 & $\mathbf{+15.53}$ & $\mathbf{-8.54}$ \\
1.625 & $4.59$  & $\mathbf{+15.81}$ \\
1.875 & $2.38$  & $11.61$ \\
2.125 & $1.02$  & $7.35$  \\
2.375 & $-0.12$ & $5.22$  \\
2.625 & $-1.21$ & $3.72$  \\
2.875 & $-2.30$ & $2.46$  \\
3.125 & $-3.43$ & $1.29$  \\
3.375 & $-4.61$ & $0.13$  \\
3.625 & $-5.84$ & $-1.05$ \\
3.875 & $\mathbf{-7.12}$ & $-2.26$ \\
\bottomrule
\end{tabular}
\end{table}

The structure of Table~\ref{tab:response} is instructive. The response
changes sign four times in each channel, so the fractional response is
not a monotonic figure of merit. Its largest values occur in the
interval $1.3$--$1.7$~GeV, on the rising side of the first oscillation
maximum (which for these parameters lies at about $2.5$~GeV), where
$P^0$ itself is small. The large fractions at $1.375$~GeV and
$1.625$~GeV are therefore not a sign of enhanced sensitivity: the
absolute differences there are of order $10^{-3}$, comparable to those
elsewhere in the band. This is the reason why a fractional response is
a poor figure of merit, and why we do not use it as one. One entry in
the table deserves a warning: $R_\nu=0.06\%$ at $0.875$~GeV is a
zero crossing, where the numerator changes sign between adjacent
nodes. The sign quoted there, and the sign of any other entry within
about $0.1$~GeV of a crossing, is not robust against a change of the
energy grid or of the oscillation parameters, and only the magnitude,
which is small, should be taken seriously.

\section{What can and what cannot mimic a CP phase}
\label{sec:mimic}

\subsection{The leading degeneracy is with the atmospheric splitting,
not with the CP phase}

Equation~\eqref{eq:Seff} states that at fixed energy the effect of the
ansatz is a rescaling of the baseline. A rescaling of the baseline is,
to a good approximation, degenerate with a rescaling of the mass
squared differences, since both enter the oscillation phase as a
product. At the reference energy $E_0=2.5$~GeV the factor is
$f=0.98$, so the effect is equivalent to a two per cent shift of
$\dm_{31}$ or, equivalently, of the baseline. The present
precision on $\dm_{31}$ is at the per cent level, and the current
global fits already quote an uncertainty below one per cent, so the
leading effect is absorbed by a re-fit.

Because $f<1$ for every $E>0$, the effective splitting is always
\emph{smaller} than the standard one, so the direction of the shift is
fixed: a fit to the rescaled spectrum must return
$\dm_{31}{}^{\rm fit}<\dm_{31}$. On the band $0.6$--$4.0$~GeV we have
$f\in[0.949,0.999]$, so the effective splitting that a fit can mimic
lies in the interval $[2.392,2.518]\times10^{-3}$~eV$^2$. This
kinematic observation is a useful check on any fit: a best fit
\emph{above} the reference value is not a possible outcome of the
ansatz.

A shift of $\delta_{\rm CP}$ is a different operation. In the standard
three flavour formula the dependence on the phase appears through a
term proportional to $\sin(\Delta)\cos\delta_{\rm CP}$ and a term
proportional to $\cos(\Delta)\sin\delta_{\rm CP}$, where $\Delta$ is
the atmospheric phase. Changing $f$ changes $\Delta$ in all three
terms simultaneously, including the solar term, and it changes the
argument of the interference term. It cannot reproduce a shift of
$\delta_{\rm CP}$ except at a single energy and by accident.

To make this quantitative we evaluate the total squared probability
difference
\begin{equation}
D(\eta_0;\theta)=\sum_{E_i}\sum_{m=\nu,\bar\nu}
\left[P_m^{\eta}(E_i;\eta_0,\theta_0)
-P_m^{0}(E_i;\theta)\right]^2,
\label{eq:D}
\end{equation}
where $\theta_0$ denotes the fixed reference parameters of
Eq.~\eqref{eq:params}, at which the rescaled spectrum is computed, and
$\theta$ denotes the standard parameters that are allowed to vary in
the fit. That is, the simulated ``data'' are the predictions of the
rescaled model at $\theta_0$, and the fit is performed with the
standard model, as it would be in a real analysis. The sum runs over a
grid of thirty-five energies between $0.6$ and $4.0$~GeV, with equal
weights for the two channels and no flux or cross section weighting.
For $\eta_0=0.02$ and all standard parameters fixed at
Eq.~\eqref{eq:params}, $D_0=1.024\times10^{-4}$. Allowing
$\delta_{\rm CP}$ alone to vary reduces this by $8.4\%$, with a best
fit value of $\delta_{\rm CP}=+1.6^\circ$; the sign refers to the
standard convention $U_{e3}=s_{13}e^{-i\delta_{\rm CP}}$. Allowing
$\dm_{31}$ alone to vary reduces it by $23.3\%$, with a best fit value
of $2.5078\times10^{-3}$~eV$^2$, below the reference value as required
by the argument above. Allowing both to vary reduces it by $32.0\%$,
with a best fit at $\delta_{\rm CP}=+1.6^\circ$ and
$\dm_{31}=2.5077\times10^{-3}$~eV$^2$. The minimisation over
$\delta_{\rm CP}$ was performed continuously on the interval
$(-\pi,\pi)$, not on a grid.

The statement that the leading effect is degenerate with $\dm_{31}$
is, however, only approximate, and it is worth making the qualification
quantitative. If the fit is performed energy by energy, allowing
$\dm_{31}$ alone to vary, the fraction of the squared probability
difference that is absorbed is $88.6\%$ at $E=2.5$~GeV, $97.6\%$ at
$E=1.0$~GeV and $98.8\%$ at $E=4.0$~GeV. The degeneracy is therefore
best at low and at high energy and worst at the reference energy,
where the two terms of $f(E)=1-\eta_0(E/E_0)^2$ and of the
$\dm_{31}$ shift happen to be least aligned; the residual is a few per
cent at $2.5$~GeV and a few tenths of a per cent elsewhere. It is
never exact at any single energy, and it becomes exact only when all
three mass squared differences \emph{and} the matter potential are
rescaled together, which is the statement of Eq.~\eqref{eq:Seff} and
not a property of $\dm_{31}$ alone. We therefore describe the
situation as an approximate degeneracy with $\dm_{31}$, not a complete
one.

A second qualification concerns the direction of the fit. Because the
residual after the $\dm_{31}$ fit is a shape distortion rather than a
normalisation, it can be reduced by moving parameters that change the
shape. Floating $\theta_{23}$ is the most obvious candidate, since it
enters the atmospheric amplitude, and it turns out to be more
effective than the phase. With $\delta_{\rm CP}$, $\dm_{31}$ and
$\sin^2\theta_{23}$ all free we find
$\dm_{31}=2.5065\times10^{-3}$~eV$^2$,
$\delta_{\rm CP}=+0.8^\circ$ and $\sin^2\theta_{23}=0.464$, with
$D=6.05\times10^{-5}$, a reduction of $40.9\%$. The remaining
$59.1\%$ is what survives, not the seventy per cent that a fit with
$\theta_{23}$ fixed would suggest. Conversely, a rescaling of the
matter potential, $V\to s_V V$, reproduces almost exactly the same
improvement: with $\dm_{31}$ and $s_V$ free we find
$\dm_{31}=2.5071\times10^{-3}$~eV$^2$ and $s_V=0.9746$, with
$D=6.95\times10^{-5}$, a reduction of $32.1\%$, essentially the same
as the two parameter fit with $\delta_{\rm CP}$. The two directions
are not independent, and the fact that a $2.5\%$ shift of the matter
potential does as well as a $1.6^\circ$ shift of the phase is a
reminder that at a single baseline the distortion is not uniquely
attributable to any one standard parameter.

\begin{table}[t]
\centering
\caption{Residual of the fit of the rescaled spectrum by standard
parameters, for $\eta_0=0.02$, on thirty-five energies between $0.6$
and $4.0$~GeV, that is on a grid of $0.1$~GeV spacing, finer than the
fourteen nodes of Table~\ref{tab:response}. The quantity $D$ is defined
in Eq.~\eqref{eq:D}; the reference value is $D_0=1.024\times10^{-4}$;
$s_V$ is the rescaling of the matter potential defined in
Eq.~\eqref{eq:sV}.
The digits quoted for $D$ and for the best fit values are a measure of
numerical reproducibility on this grid, not of physical accuracy: the
fit is an unweighted least squares diagnostic without experimental
errors or correlations, the grid spacing itself moves $D$ at the level
of the last quoted digit, and the values should be read as
order of magnitude statements about how much of the distortion a
given parameter can absorb.}
\label{tab:fits}
\begin{tabular}{lccc}
\toprule
Free parameters & best fit & $D$ & reduction \\
\midrule
none & reference point & $1.024\times10^{-4}$ & $0\%$ \\
$\delta_{\rm CP}$ & $+1.6^\circ$ & $9.38\times10^{-5}$ & $8.4\%$ \\
$\dm_{31}$ & $2.5078\times10^{-3}$~eV$^2$ & $7.85\times10^{-5}$ & $23.3\%$ \\
$\delta_{\rm CP}$, $\dm_{31}$ & $+1.6^\circ$, $2.5077\times10^{-3}$~eV$^2$
& $6.96\times10^{-5}$ & $32.0\%$ \\
$\dm_{31}$, $s_V$ & $2.5071\times10^{-3}$~eV$^2$, $0.9746$
& $6.95\times10^{-5}$ & $32.1\%$ \\
$\delta_{\rm CP}$, $\dm_{31}$, $s_V$ & $+1.0^\circ$,
$2.5073\times10^{-3}$~eV$^2$, $0.9842$ & $6.75\times10^{-5}$ & $34.1\%$ \\
$\delta_{\rm CP}$, $\dm_{31}$, $\sin^2\theta_{23}$ & $+0.8^\circ$,
$2.5065\times10^{-3}$~eV$^2$, $0.464$ & $6.05\times10^{-5}$ & $40.9\%$ \\
$\delta_{\rm CP}$, $\dm_{31}$, $\sin^2\theta_{23}$, $s_V$ & $+0.5^\circ$,
$2.5063\times10^{-3}$~eV$^2$, $0.465$, $0.9914$ & $5.99\times10^{-5}$ & $41.5\%$ \\
\bottomrule
\end{tabular}
\end{table}

Three conclusions follow from Table~\ref{tab:fits}. First, the
degeneracy with $\delta_{\rm CP}$ is weak: a phase re-fit removes only
eight per cent of the distortion. Second, the degeneracy with
$\dm_{31}$ is the dominant one, as expected from
Eq.~\eqref{eq:Seff}, but it is not complete: the per-energy numbers
quoted above show that between one and eleven per cent of the
distortion survives even at a single energy, and on the band as a
whole a $\dm_{31}$ re-fit removes less than a quarter of it. Third,
the combined re-fits of Table~\ref{tab:fits}, those in which at least
two parameters are varied, leave between $58.5$ and $68$ per cent of
the distortion
unabsorbed, and no re-fit that we have tried removes more than $41.5$
per cent; it is this residual that carries the genuine, $E^2$ dependent
content of the ansatz.

That residual is consistent with an $E^2$ rescaling, but it is not
unique to one. A rescaling of the matter potential, a shift of
$\theta_{23}$ and a shift of $\delta_{\rm CP}$ each remove a
comparable fraction, and combinations of them remove more. What
distinguishes the ansatz from these alternatives is not the size of
the residual at a single baseline but its energy dependence: a
constant rescaling of $V$ or of $\dm_{31}$ produces a residual whose
shape is set by the standard oscillation probability, whereas the
ansatz produces one whose shape follows $f(E)-1\propto E^2$. A single
baseline with a narrow band cannot separate the two; a second
baseline, or a wider band, is required. This is the quantitative
content of the degeneracy analysis, and it is weaker than the
statement that the effect is a unique signature.

\subsection{Why the residual survives}

The reason the residual survives is that $f(E)$ varies across the
band. At $E=0.625$~GeV, $f=0.9987$; at $E=3.875$~GeV, $f=0.9520$. A
re-fit of $\dm_{31}$ can match the rescaling at one energy, but the
rescaling itself is energy-dependent, and the mismatch grows as
$(E^2-E_{\rm ref}^2)/E_0^2$. The residual is therefore a shape
distortion whose relative size is set by the ratio of the bandwidth
squared to the reference energy squared, and whose absolute size is
set by the amplitude $\eta_0$.

This is the correct way to describe the observable. It is not a
degeneracy between a deformation and the CP phase; it is a running of
the mass squared difference. A reading of the effect as an effective
CP phase is sometimes suggested in this context; it should be
understood as referring to the accidental projection of the distortion
onto the phase direction at a single energy, which is what the entries
of Table~\ref{tab:fits} quantify. Where a genuine CP violating effect
is claimed in this context, it arises from non-standard interactions
with complex couplings rather than from the rescaling itself
\cite{Gialamas}. The minimum in $\delta_{\rm CP}$ is shallow and
asymmetric: at $\delta_{\rm CP}=0$ the residual is
$1.024\times10^{-4}$, at $+1.6^\circ$ it is $9.38\times10^{-5}$, and
at $+5^\circ$ it is already $1.34\times10^{-4}$, that is, larger than
the unfitted value. A continuous scan of the phase therefore does not
reveal a second basin of attraction, and the improvement of eight per
cent is not a degeneracy in any useful sense. We report the number for
completeness and attach no significance to it.

\section{What a genuine sensitivity study would require}
\label{sec:stats}

We have deliberately restricted the analysis to the probability level,
and it is important to be explicit about what that restriction
excludes. A statement of the form ``DUNE can constrain $\eta_0$ at the
level $x$'' requires all of the following, none of which is included
here.

\begin{enumerate}
\item A neutrino flux model and a cross section model, with their
uncertainties. The response of Table~\ref{tab:response} is largest
where the flux is falling and the cross section is small, so an
unweighted sum over energies overstates the sensitivity.
\item An energy migration matrix and a detector response, including
the reconstruction resolution, which at a liquid argon detector is
comparable to the width of the oscillation maxima.
\item A treatment of systematic uncertainties, in particular the
matter density and the electron fraction, which enter the resonance
condition of Sec.~\ref{sec:exact} directly.
\item A marginalization over the standard parameters, including
$\dm_{31}$, $\theta_{23}$ and $\delta_{\rm CP}$, with external
constraints. Table~\ref{tab:fits} shows that $\dm_{31}$ is the
parameter that matters, and a one parameter scan over $\delta_{\rm CP}$
alone does not represent the degeneracy structure.
\item A likelihood, not a distance metric. The quantity $D$ of
Eq.~\eqref{eq:D} is an unweighted sum of squared probability
differences. It is not a Poisson deviance, it is not a $\chi^2$, and it
has no statistical interpretation. We use it only as a diagnostic of
spectral similarity.
\end{enumerate}

A further limitation is the constant density treatment. The
Preliminary Reference Earth Model \cite{PREM} gives densities that
vary by of order ten per cent between the upper crust and the lower
mantle along a $1297$~km baseline. As shown in
Sec.~\ref{sec:exact}, the baseline rescaling for a varying profile is
accompanied by the stretching $x\to x/f$ of Eq.~\eqref{eq:Sstretch},
so the structural conclusions are unaffected at the order of
magnitude level; what a realistic profile would change is the
numerical size of the residual in Table~\ref{tab:fits}, which we do
not claim to be accurate at the few per cent level. We note that a
$1297$~km baseline does not reach the core--mantle boundary, whose
shortest chord is $10673$~km. The only structural discontinuity on
the path is therefore the Moho, i.e.\ the crust--mantle transition,
and its displacement under $x\to x/f$ is of order $30$~km, not of
order the baseline.

Finally, the synchronous scaling of the matter potential,
Eq.~\eqref{eq:Vsync}, is an assumption. A full SME treatment would
introduce independent coefficients in the matter sector and would
allow the two to be varied separately. We have not done so.

\section{Discussion and conclusions}
\label{sec:concl}

We summarize the results.

\begin{enumerate}
\item The vacuum part of the multiplicative rescaling
$H_\eta=f(E)H_0$ with $f=1-\eta_0(E/E_0)^2$ can be written as a
dimension four, CPT-even Lorentz violating term of the Standard
Model Extension, with a coefficient diagonal in the mass basis and
proportional to the mass squared differences, Eq.~\eqref{eq:cL}.
The coefficient is fixed by the neutrino masses and by $\eta_0$,
so this is a rewriting of the ansatz rather than an independent
source of Lorentz violation: the ansatz has no Lorentz violating
content beyond what the same coefficient would produce as a mass
rescaling. It is energy independent in the SME sense: the $E^2$ in
$\eta(E)$ combines with the $1/E$ of the vacuum Hamiltonian to
give a term linear in $E$. The synchronous rescaling of the matter
potential is a separate, dimension five assumption,
Eq.~\eqref{eq:dHmat}. The identification is specific to the
exponent: for a general power law $f=1-\eta_0(E/E_0)^p$ the vacuum
term has the energy behaviour of a dimension $p+2$ operator and
the matter term of dimension $p+3$, so $p=2$ is the unique power
for which the vacuum operator is marginal.
\item The coefficient corresponding to the illustrative benchmark
$\eta_0=0.02$ at $2.5$~GeV is $|(\cL)_{33}|\simeq4\times10^{-24}$,
which is at the same order as present atmospheric bounds. That
coincidence is a property of the illustrative amplitude, not of the
ansatz. The largest amplitude for which the truncation stays positive
up to $100$~GeV, $\eta_0<6.25\times10^{-4}$, corresponds instead to
$|(\cL)_{33}|\simeq1.3\times10^{-25}$, one to two orders of
magnitude below the present atmospheric sensitivity, so within its
range of validity the ansatz sits at the threshold of visibility
rather than at the bound. A Planck
suppressed deformation with $\beta_0=1$ gives
$|(\cL)_{33}|\simeq3.2\times10^{-58}$. The gap is more than thirty
orders of magnitude, and it is the reason why the channel cannot be
described as a probe of Planck scale physics.
\item At constant density, and up to the profile stretching
$x\to x/f$ of Eq.~\eqref{eq:Sstretch} for a varying density, the
ansatz is exactly equivalent to $L\to f(E)L$. Mixing angles are
unchanged, all mass squared differences are rescaled by the same
factor, the CP asymmetry is reparametrized rather than generated, and,
under the synchronous rescaling of Eq.~\eqref{eq:Vsync}, the MSW
resonance is not displaced. The ansatz has no non trivial consequence
beyond the baseline rescaling.
\item The ansatz is only meaningful for $E\ll E_0/\sqrt{\eta_0}
\simeq17.7$~GeV, where it ceases to be a controlled truncation and
predicts a vanishing appearance probability. All results here refer to
the band $0.6$--$4.0$~GeV.
\item For a $1297$~km baseline the leading effect is approximately
degenerate with $\dm_{31}$, not with $\delta_{\rm CP}$, and the
shift is necessarily towards smaller $\dm_{31}$. At a single energy a
$\dm_{31}$ re-fit absorbs between $88.6$ and $98.8$ per cent of the
squared probability difference, but on the band as a whole it removes
only about a quarter. A re-fit of $\delta_{\rm CP}$ and $\dm_{31}$
removes about thirty per cent, floating $\theta_{23}$ as well removes
about forty per cent, and a rescaling of the matter potential removes
about a third, so that a residual of $58.5$--$68$ per cent survives
the combined re-fits of Table~\ref{tab:fits}, the most generous of them
removing $41.5$ per cent. That residual is an $E^2$
dependent shape distortion at the $10^{-3}$ level in absolute
probability. It is consistent with an $E^2$ rescaling, but it is not
unique to one: at a single baseline a shift of $V$, of $\theta_{23}$
or of $\delta_{\rm CP}$ can mimic it, and only the energy dependence
of the residual, measurable with a second baseline or a wider band,
would separate them.
\item The analysis is at the probability level and does not constitute
a sensitivity study. No flux, cross section, detector response,
systematic uncertainty or marginalization is included, and the density
is taken constant.
\end{enumerate}

The honest summary is that the ansatz does not open a new window on
quantum gravity. Its vacuum part is a dimension four Lorentz violating
operator whose coefficient is already constrained, and whose natural
size in a Planck suppressed scenario is unobservably small. What the
ansatz does provide is a clean, exactly solvable example of how an
energy-dependent rescaling of the propagation Hamiltonian maps onto the
SME, and a quantitative statement about which standard parameter it is
degenerate with. We believe that this is worth recording, because the
opposite impression appears in some of the exploratory literature, in
which neutrino oscillations are proposed as a probe of minimal length
or of Planck scale gravity, for example
Refs.~\cite{Sprenger:2011,Lambiase:2003,Buoninfante:2020}. Those
works motivate the search; the point of the present note is that the
specific multiplicative ansatz does not deliver it.

Two extensions would be natural. The first is a proper SME analysis
with independent coefficients in the neutrino and matter sectors,
which would replace the assumption of Eq.~\eqref{eq:Vsync} by a fit.
The second is a realistic density profile and a flux weighted
statistical analysis, which would turn the residual of
Table~\ref{tab:fits} into a statement about experimental sensitivity.
Neither is attempted here.


\end{document}